\documentclass[12pt]{article}
\begin{document}
\title{STUDY OF PROTO STRANGE STARS (PSS) IN TEMPERATURE AND DENSITY DEPENDENT QUARK MASS MODEL}
\author{V.K.Gupta \footnote{E--mail : vkg@ducos.ernet.in}, Asha Gupta, S.Singh\footnote{E--mail : santokh@ducos.ernet.in}, J.D.Anand\footnote{E--mail : jda@ducos.ernet.in} \\
 	{\em Department of Physics and Astrophysics,} \\
	{\em University of Delhi, Delhi-110 007, India.} \\
        {\em InterUniversity Centre for Astronomy and Astrophysics,} \\
        {\em Ganeshkhind, Pune 411007 , India.} \\
        }
\renewcommand{\today}{}
\setlength\textwidth{5.75 in}
\setlength\topmargin{-1.cm}
\setlength\textheight{8 in}
\addtolength\evensidemargin{-1.cm}
\addtolength\oddsidemargin{-1.cm}
\font\tenrm=cmr10
\def\baselinestretch{1.4}

\maketitle
\large
\begin{abstract}
We report on the study of the mass-radius (M-R) relation and the radial oscillations of proto strange stars. For the quark matter we have employed the well known density dependent quark mass model and its very recent modification, the temperature and density dependent quark mass model. We find that the maximum mass the star can support increases significantly with the temperature of the star in this model which implies that transition to a black hole at the early stage of formation of the star is inhibited. As for the neutrinos, we find, contrary to the expectation that the M-R and oscillation frequencies are almost independent of the neutrino chemical potentials.\\  
Subject headings: Strange stars - Oscillations
\end{abstract}
\pagebreak

\begin{section} {Introduction}

It is believed at present that at high densities ($\rho\geq 3\rho_0$) the hadronic matter undergoes a phase transition to the unconfined state of quarks and gluons which is composed of roughly equal number of u, d and s quarks together with a gas of electrons and muons necessary to maintain charge neutrality. This strange quark matter (SQM) may be the true ground state of matter (Witten 1984; Farhi $\&$ Jaffe 1984; see also e.g. Madsen 1999, for a recent review of the physics and astrophysics of strange quark matter and strange stars). Since then, various implications of SQM have been explored in astrophysics and high energy physics. Glendenning (1997) has conjectured the nucleation of SQM bubbles and most of the currently believed neutron stars may be strange stars.\\
Pulsars can be modeled as neutron stars or strange stars. It is very difficult to distinguish observationally between the two of them. Recently three ways have been proposed for identifying a strange star:\\
1. A hot strange star or a proto strange star (PSS) may rotate more rapidly since its higher bulk viscosity (Wang and Lu 1984; Madsen 1992; Anand et al. 2000a) can effectively damp away r-mode instability (Madsen 1998).\\
2. Mass-Radius (M-R) relations of strange stars (M $\propto$ $R^3$) is quite different from that of neutron stars (Li et al. 1999).\\
3. It is possible to distinguish "bare" polar strange stars from neutron stars via pulsar magnetospheric and polar radiation (Xu et al. 1999; Xu et al. 2001).
A newly born star, be it a neutron star or strange star passes through various stages of evolution. The duration of these stages is essentially determined by neutrino diffusion time scales (Prakash et al. 1997). Roughly the diffusion time scale is proportional to $R^2\lambda$ where R is the
 star radius and $\lambda$ the neutrino mean free path. Both the M-R relation and mean free path depend upon the equation of state (EOS) and the composition of the star matter. The behaviour of the maximum mass as a function of the temperature and neutrino trapping is also of practical importance to the process of conversion of the star into a black hole and the stage at which this transition occurs, if at all.\\ 
 The formation of proto neutron stars (PNS) is relatively well understood (for a recent review see e.g. Prakash 1997; Glendenning 1997). But due to the absence of a full theory  determining the conditions at which quark matter phase transition occurs and a lack of detailed understanding of the complex burning process of neutron matter into strange matter, very little is known about how a PNS converts to PSS during the supernova explosion. However some studies of the transition of nuclear matter to SQM taking into account the two step phase transition viz neutron matter $\rightarrow$ ud $\rightarrow$ uds have been carried out by numerous authors (Dai et al. 1995; Anand et al. 1997; Lugones and Benvenuto 1998,1999).\\
In the first step the neutron matter deconfinement occurs on a stronge interactions time scale $\sim$ $10^{-23}$s; after that chemical equilibrium of the deconfined matter takes place on a weak interaction time scale $\sim$ $10^{-8}$s. Additional neutrinos and energy are produced in the second step. Recently Gupta et al. (2002) have studied the phase transition in a PNS from nuclear matter to two flavour quark matter.\\
Most of the calculations for the phase transition and the properties of strange stars have been done in the MIT Bag model. There is an alternative and equally attractive approach based on the density dependent quark mass (DDQM) model originally proposed by Fowler, Raha $\&$ Weiner (1981). Many aspects of SQM like radial oscillations of both rotating and non rotating quark stars, bulk viscosity etc. have been studied by various authors in this model (Anand et al. 2000a; Anand at al. 2000b). At zero and very low temperatures the model works reasonably well and gives results for various quantities like the M-R relation and viscosity etc. which are qualitatively similar to those obtained with bag model. Very recently Zhang $\&$ Su (2002) have pointed out that the DDQM model suffers from a basic drawback that it can not be used to explain the process of quark deconfinement phase transition as the quark confinement is permanent in this model. They have suggested a temperature and density dependent quark mass (TDDQM) model to describe the bulk SQM. In this model the mass of quarks not only depends on baryon density (as in DDQM model) but also on the temperature. In DDQM model the masses of u, d and s quarks are taken to be (Fowler, Raha $\&$ Weiner 1981; Lugones and Benvenuto 1998)
\begin{eqnarray}
    m_u = m_d =\frac{C}{3n_B},  \nonumber\\
    m_s = m_{so} + \frac{C}{3n_B}
\end{eqnarray}
In TDDQM model guided by Friedman-Lee soliton bag model (Lee 1981), C depends on temperature and can be parameterized as
\begin{eqnarray}
      C(T) &=& C_o~ [1-({\frac{T}{T_c}}^2]; ~~~~0~\leq~T~\leq~T_c \nonumber\\
          &=& 0;~~~~~~~~~~~~~~~~~~~~~~~~~~~T~>~T_c   
\end{eqnarray}     
where $T_c$ is the critical temperature of deconfinement. At T = 0, it of course reduces to the usual DDQM model. One system where the results can be expected to be affected significantly by the inclusion of this temperature dependence of the quark masses is the PSS which are born with a rather high temperature of $\sim$ 60 MeV or more and large lepton excess. Motivated by these factors we have studied in this paper the M-R relations and radial oscillations of PSS during the first few ($\sim 20)$ seconds of its existence when it has a high temperature and large number of electron (and possibly muon) neutrinos. In section 2, we present the formalism and the thermodynamics used. Section 3 deals with M-R relation and radial oscillations of PSS and section 4 is devoted to results and discussion.
\end{section}

\vspace {0.5cm}
\begin{section} {Formalism and Thermodynamics}

We consider PSS as a system of uds quarks along with electron, muon and their neutrinos. The thermodynamical potential for the system is given by $\Omega = \sum\Omega_i$, where
\begin{eqnarray}
\Omega_i &=& - \frac{g_iT}{2\pi^2} \int dp[\log(1+\exp(-\beta[E_i -\mu_i])) \nonumber \\
 &+&\log(1+\exp(-\beta[E_i +\mu_i]))]
\end{eqnarray}
and 
\begin{equation}
E_i = \sqrt{p^2 +m_i^2}
\end{equation}
$g_i$= 6 (3 colour $\times$ 2 spin) for u, d and s quark, two for e and $\mu$ and one for $\mu_{\nu_e}$ and $\mu_{\nu_\mu}$. For quarks, the expressions for the masses in the TDDQM model are given by equation (1) with C replaced by C(T) of equation (2). The thermodynamic quantities of interest like the energy density, $\rho$, the pressure, P and various number densities $n_i$ can be obtained from the thermodynamical potential $\Omega_i$:
\begin{equation}
P_i = -\frac{\partial(\Omega_{i}/n_{B})}{\partial (1/n_{B})} \nonumber\\
= n_{B} \frac{\partial\Omega_{i}}{\partial n_{B}} -\Omega_{i} \nonumber\\
= n_{B} \frac{\partial\Omega_{i}}{\partial m_{i}} \frac{\partial m_{i}}{\partial n_{B}}  -\Omega_{i} 
\end{equation}
The additional term $ n_B \frac{\partial\Omega_i}{\partial n_B}$ arises because of the dependence of the quark masses on the baryon density.
\begin{equation}
\rho_i =\Omega_i +\mu_in_i -T \frac{\partial\Omega_i}{\partial T} 
\end{equation}
and number density $n_i$ by
\begin{equation}
n_i = - \frac{\partial\Omega_i}{\partial\mu_i} 
\end{equation}
Using eqn (3) for the thermodynamical potential we get, for the quarks
\begin{eqnarray}
P_i &=&\frac{1}{3} \frac{g_i}{2\pi^2}\int\frac{p^4dp[f_{i}(T)+\bar{f_i}(T)]}{E_{i}}\nonumber \\
&-&\frac{C}{3 n_{B}} \frac{g_i m_{i}}{2\pi^2}
\int\frac{p^2dp[f_{i}(T)+\bar{f_i}(T)]}{E_{i}}
\end{eqnarray}
\begin{eqnarray}
\rho_i &=& \frac{g_i}{2\pi^2} \int p^2E_i dp(f_{i}(T)+\bar{f_i}(T))\nonumber\\
&+& \frac{1}{3}\frac{g_i m_{i}C_o T^2}{\pi^2 n_{B}T_{c}^2}
\int\frac{p^2dp[f_{i}(T)+\bar{f_i}(T)]}{E_{i}}
\end{eqnarray}
\begin{equation}
n_i = \frac{g_i}{2\pi^2} \int p^2dp[f_{i}(T)-\bar{f_i}(T)]
\end{equation}
Here $f_i(T)$ and $\bar{f_i}(T)$ are the fermi-dirac distribution functions for particles and antiparticles respectively:
\begin{equation}
f_i(T) = (\exp([\sqrt{p^2+m_i^{2}} - \mu_i]/T)+1)^{-1}
\end{equation}
\begin{equation}
\bar{f_i}(T) = (\exp([\sqrt{p^2+m_i^{2}} + \mu_i]/T)+1)^{-1}
\end{equation}
The pressure and energy of electrons, muons and their neutrinos is given by the same expressions (8) and (9), except that the second term does not contribute. The total Pressure, P and energy density, $\rho$ are the sum of contributions from each species of particles: 
\begin{eqnarray}
P &=& \Sigma P_i + \frac{8}{45}\pi{^2}T^4
\end{eqnarray}
\begin{eqnarray}
\rho &=& \Sigma \rho_i+ \frac{8}{15}\pi{^2}T^4
\end{eqnarray}  
where the last term is the contribution of the gluons. \\
Having derived in detail the thermodynamics with variable particle masses, we now apply it to the investigation of PSS. Since PSS is a  mixture of u, d, s quarks, electrons, muons and their neutrinos with $\beta$ equilibrium maintained by the weak interactions,\\
$~~~~~~~d,s~\leftrightarrow~u+e+\bar{\nu_e},~~~~s+u~\leftrightarrow~u+d$\\
$~~~~~~~~~~~~\leftrightarrow~u+\mu+\bar{\nu_{\mu}}$\\
the various chemical potentials $\mu_i$ (i =u, d, s, e, $\nu_e$ and $\nu_\mu$) satisfy
\begin{eqnarray}
\mu_d=\mu_s, ~~~ \mu_d = \mu_u +\mu_e - \mu_{\nu_e}\nonumber\\
\mu_\mu-\mu_{\nu_\mu} = \mu_e - \mu_{\nu_e}
\end{eqnarray}
Charge neutrality of the system demands
\begin{equation}
\frac{2}{3}n_u-\frac{1}{3}n_d-\frac{1}{3}n_s-n_e-n_\mu=0
\end{equation}
The total baryon density is given by
\begin{equation}
n_B=\frac{1}{3}(n_u+n_d+n_s) 
\end{equation}
Equations (15 - 17) can be solved self consistently for any given baryon density, temperature and the two neutrino chemical potentials. Once these equations have been solved the pressure, P, and energy density, $\rho$, can be obtained from equation (13) and (14) and these give the required EOS, P($\rho$).
\end{section}

\vspace {0.5cm}
\begin{section} {Radial Pulsations of PSS}
The radial pulsations of a non rotating star in general relativity were first studied by Chandrasekhar (1964). The metric used is 
\begin{equation}
d{s^2}=-e^{\nu}c^{2}dt^{2}+e^{\lambda}dr^{2}+r^{2}(d{\theta}^{2}+sin^{2}{\theta}d\phi^{2})
\end{equation}
In hydrostatic equilibrium, the structure of the star is described by the 
Tolman-Openheimer-Volkoff equations
\begin{equation}
\frac{dm}{dr}=4\pi{r^{2}}\rho
\end{equation}
\begin{equation}
\frac{dP}{dr}=\frac{-G(P+\rho c^2)(m+4\pi r^3Pc^{-2})}{c^2r^2(1-\frac{2GM}{c^2r})}
\end{equation}
\begin{equation}
\frac{d\nu}{dr}=\frac{2GM(1+4\pi{r^3}Pmc^{-2})}{c^2r(1-\frac{2GM}{c^2r})}
\end{equation}
For given equation of state P($\rho$), equations (19-21) can be numerically integrated outwards starting from r = 0 with a given central density $\rho_c$ and the corresponding pressure $P_c(\rho_c)$. The equations are solved upto the point where P = 0 which defines the radius, R, of the star. The value of m(R) = M is the gravitational mass of the star.\\
We next look at the radial oscillations of the PNS. If ${\delta}r$ is the radial displacement,
\begin{equation}
\xi=\frac{{\delta}r}{r} ,~~~ \zeta=r^{2}e^{-\nu/2}\xi
\end{equation} 
and projecting the time dependence as $\exp(i\sigma t)$, one gets the equation governing radial adiabatic oscillations (Chandrasekhar 1964; Datta et al 1998 and Anand et al 2000a)
\begin{equation}
F\frac{d^{2}\zeta}{dr^{2}}+G\frac{d\zeta}{dr}+H\zeta=\sigma^{2}\zeta
\end{equation}
where
\begin{equation}
F=-\frac{e^{\nu-\lambda}}{P+\rho c^2}(\Gamma P)
\end{equation}
\begin{equation}
G=-\frac{e^{{\nu}-{\lambda}}}{P+\rho c^2}\Bigg[\frac{1}{2}{\Gamma}P(\lambda+3\nu)+\frac{d({\Gamma}P)}{dr}-\frac{2}{r}({\Gamma}P)\Bigg]
\end{equation}
\begin{equation}
H=\frac{e^{{\nu}-{\lambda}}}{P+\rho c^2}\Bigg[\frac{4}{r}\frac{dP}{dr}+\frac{8{\pi}Ge^{\lambda}P(P+\rho c^2)}{c^4}-\frac{1}{P+\rho c^2}(\frac{dP}{dr})^{2}\Bigg]
\end{equation}
$\lambda$ is related to the metric function through
\begin{equation}
e^{-\lambda}=(1-\frac{2GM(r)}{r c^2})
\end{equation}
and
\begin{equation}
\Gamma P=\frac{P+\rho c^2}{c^2}\frac{dP}{d\rho}
\end{equation}
 The boundary conditions to solve the pulsations equations (23)-(28) are
\begin{equation}
\zeta(r=0) = 0 
\end{equation}
\begin{equation}
{\delta}P(r=R) = 0
\end{equation}
Equation (23) subject to boundary conditions (29) and (30) represent a Sturm-Liouville eigenvalue problem for $\sigma^{2}$. The radial oscillations equations (23) to (26) are totally model independent and are infact the same whether one considers a neutron star, quark star, PNS or PSS. The nature of radial oscillations depends only on the equation of state used.
\end{section}

\vspace {0.5cm}
\begin{section} { Results and discussion }
In the TDDQM model there are three free parameters $C_o$, $m_{so}$ and $T_c$. $C_o$ and $m_{so}$ are fixed by the requirement that at T = 0 uds should be absolutely stable, i.e. Energy/baryon $\leq$ 930 MeV while for ud, E/$n_B$ $>$ 940 MeV. This gives us a stability window for $C_o$ and $m_{so}$. We have chosen two sets, (i) $C_o$ = 185 MeV$fm^{-3}$ and $m_{so}$ = 150 MeV, and (ii) $C_o$ = 210 MeV$fm^{-3}$ and $m_{so}$ = 100 MeV, which both lie within the stability window. As explained in the introduction, following Zhang and Su (2002) we have chosen $T_c$ = 170 MeV. The chemical potentials are evaluated self consistently for a given baryon density, temperature and neutrino chemical potentials by demanding charge neutrality and $\beta$ equilibrium conditions for a given set of values of $C_o$ and $m_{so}$. This provide us a profile of pressure P and energy density $\rho$. Using the EOS i.e. $\rho$, P profile, we then solve the stellar structure equations that provide PSS configurations. The numerical integration was done by using Runge-Kutta integration procedure to obtain p,m and $\nu$ as functions of r. We then calculate ($\Gamma$P) directly from the EOS for all densities by using quadratic difference formula for derivative ($\frac{dP}{dr}$). The equations (23)-(28) were also solved by Runge-Kutta integration method. By changing the number of input points, it was estimated that $\sigma^2$ is accurate to one part in $10^3$.\\
In fig. 1, we plot the gravitational mass M in terms of solar mass vs radius R in Km for $\mu_{\nu_e}$ = $\mu_{\nu_\mu}$ = 0 MeV. Curves a, b, c, and d correspond to the temperatures T = 0, 20, 40 and 60 MeV respectively in fig. 1 to fig. 4. It is seen that the maximum mass and the radius increases with the temperature. The curve a) for T = 0 corresponds to the normal strange star which has been studied in the same model by Benvenuto and Lugones (1998) and Anand et al. (2000a). The present results are different from those obtained by the above authors, since, as has been pointed out by Peng et al. (2000) and Zhang $\&$ Su (2002), the expression for the energy density used by these authors is different and inconsistent in that the lowest energy state does not correspond to zero pressure. Thus the present results must be considered more realistic in the DDQM or the TDDQM models.\\ 
Fig.2 to fig.4 show the graphs of M vs. R for three other neutrino chemical potentials ($\mu_{\nu_e}$, $\mu_{\nu_\mu}$): (0,200); (200,0) and (200,200) MeV respectively. The general features are the same as those in fig.1. It seems that M-R relations are not too sensitive to the neutrino chemical potentials, as against PNS whose structure is expected to depend quite significantly on the presence of neutrinos (Prakash et al. 1997). For this reason for the rest of the investigations we consider $\mu_{\nu_e}$ = $\mu_{\nu_\mu}$ = 200 MeV only.\\
Fig.5 deals with the variation of maximum mass with the temperature. We find that the maximum mass increases significantly with increase in temperature implying that PSS supports higher mass when newly born. This, according to Prakash et al. (2000) implies that transition to black hole can not take place very early; if it takes place at all, it could happen later on thermal diffusion time scale $\sim$ 10 s. In fig.6 we show the effect of the strange quark mass on our calculation. We chose two sets of $C_o$ and $m_{so}$  (185, 150); (210, 100) which both correspond to the same minimum energy per baryon, viz $\sim$ 924 MeV. We see that the maximum mass of the PSS vary by about 10$\%$ over this range of strange quark masses. In this figure we also show the effect of inclusion of the temperature dependence of the quark masses. Curve a) corresponds to the TDDQM model whereas curve c) corresponds to DDQM model. Inclusion of the temperature dependence of the quark masses increases the maximum mass as well as the radius by about 5-10$\%$.\\
In fig.7, we plot frequency of radial pulsations for fundamental (n = 0) and first excited (n = 1) mode vs mass of PSS for $\mu_{\nu_e}$ = $\mu_{\nu_\mu}$ = 200 MeV and temperatures T = 0 and 60 MeV. We notice that at any temperature the fundamental mode frequency is exactly equal to zero for maximum mass of the star for that temperature implying that the numerical code used by us is in order. At higher temperatures the frequencies of cold and very hot PSS are almost same implying that the hotness of the star does not influence much the radial oscillations of the star. In fig.8 we show the effect of strange quark mass and temperature dependence on the radial frequencies for n=0 and n=1 modes. Curves a) and c) correspond to the two sets of $C_o$ and $m_{so}$  (185, 150); (210, 100) respectively for n=0 and (d,f) for n=1. We notice that the effect is rather small but relatively more pronounced for n=1. Same is true for the effect of temperature dependence of the quark mass which can be obtained by comparision of curves (a, b) for n=0 and (d, e) for n=1.\\
\\To conclude, we find that the correct thermodynamic treatment (Peng et al. 2000) of density dependent masses changes the results significantly. The PSS being at a higher temperature comparred to normal SS has a much higher maximum mass and can thus support a star of higher mass, implying that transition to black hole at a very early stage is inhibited. However, contrary to the usual expectation neutrino chemical potentials do not seem to have much effect on M-R and the radial pulsations of the PNS. Thus the structure depends upon the temperature with which the PSS is born and not on the lepton fraction.
Since the magnetic field has a considerable effect on the possibility of phase transition (Gupta et al. 2002), it will be interesting to study its effect on the structure of PSS. This work is in progress and will be reported elsewhere.
\end{section}
\pagebreak 

Figure captions
\vskip 0.5 cm
Figure 1. Radius R in km vs. mass M/$M_{\odot}$ (where $M_{\odot}$ is the solar mass) for $C_o$ = 185 MeV$fm^{-3}$, $m_{so}$ = 150 MeV, $T_c$ = 170 MeV and $\mu_{\nu_e}$ = $\mu_{\nu_\mu}$ = 0 MeV. The curves a, b, c and d correspond to T = 0, 20, 40 and 60 MeV respectively.    
\vskip 0.5 cm
Figure 2. Same as fig. 1 but for $\mu_{\nu_e}$ = 0 MeV and $\mu_{\nu_\mu}$ = 200 MeV. 
\vskip 0.5 cm
Figure 3. Same as fig. 1 but for $\mu_{\nu_e}$ = 200 MeV and $\mu_{\nu_\mu}$ = 0 MeV.
\vskip 0.5 cm
Figure 4. Same as fig. 1 but for $\mu_{\nu_e}$ = $\mu_{\nu_\mu}$ = 200 MeV.
\vskip 0.5 cm
Figure 5. Temperature T vs. $M_{max}/M_{\odot}$ (where $M_{max}$ is the maximum mass) for $C_o$ = 185 MeV$fm^{-3}$, $m_{so}$ = 150 MeV, $T_c$ = 170 MeV and $\mu_{\nu_e}$ = $\mu_{\nu_\mu}$ = 200 MeV.
\vskip 0.5 cm
Figure 6. Radius R vs. M/$M_{\odot}$ for $\mu_{\nu_e}$ = $\mu_{\nu_\mu}$ = 200 MeV and T = 60 MeV. Curves a, b correspond to $T_c$ = 170 MeV and two sets of $C_o$ $\&$ $m_{so}$ (185, 150); (210, 100) respectively. Curve c corresponds to DDQM model at C = 185 MeV$fm^{-3}$ and $m_{so}$ = 150 MeV. 
\vskip 0.5 cm
Figure 7.  Mass M/$M_{\odot}$ vs. frequency $\nu$ (KHz) for $C_o$ = 185 MeV$fm^{-3}$, $m_{so}$ = 150 MeV, $T_c$ = 170 MeV and $\mu_{\nu_e}$ = $\mu_{\nu_\mu}$ = 200 MeV. Curves a, b correspond to fundamental mode, n = 0, at T = 0 and 60 MeV respectively. Curves c, d correspond to n = 1 mode at T = 0 and 60 MeV respectively. 
\vskip 0.5 cm
Figure 8. Mass M/$M_{\odot}$ vs. frequency $\nu$ (KHz) for $\mu_{\nu_e}$ = $\mu_{\nu_\mu}$ = 200 MeV and T = 60 MeV. Curves a, c correspond to $T_c$ = 170 MeV and two sets of $C_o$ $\&$ $m_{so}$ (185, 150); (210, 100) respectively for fundamental mode. Curves b, e corresponds to corresponding DDQM model at C = 185 MeV$fm^{-3}$ and $m_{so}$ = 150 MeV. 

\pagebreak

\begin{figure}[ht]
\vskip 15truecm
\includegraphics{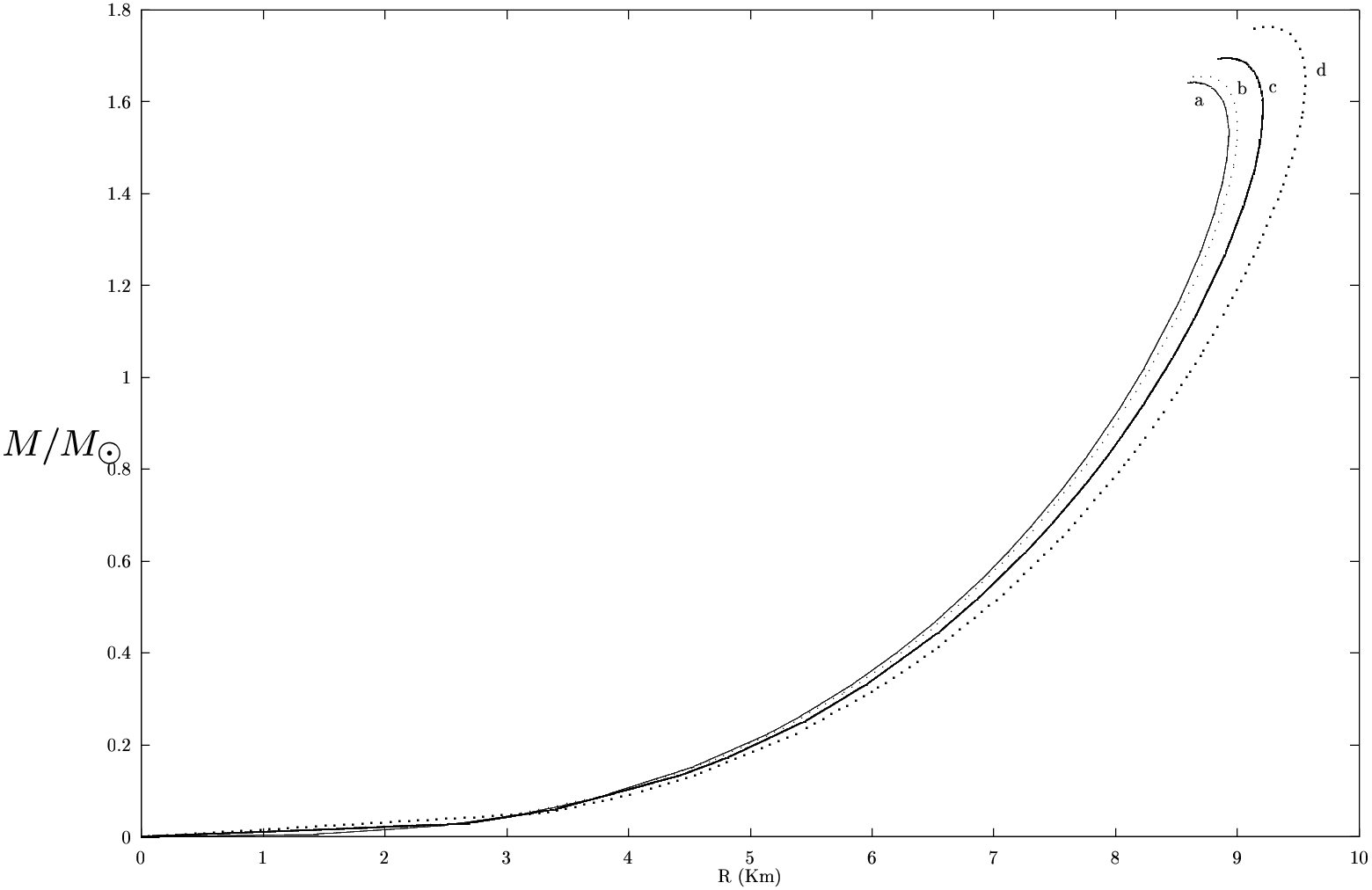}
\caption{Radius R in km vs. mass M/$M_{\odot}$ (where $M_{\odot}$ is the solar mass) for $C_o$ = 185 MeV$fm^{-3}$, $m_{so}$ = 150 MeV, $T_c$ = 170 MeV and $\mu_{\nu_e}$ = $\mu_{\nu_\mu}$ = 0 MeV. The curves a, b, c and d correspond to T = 0, 20, 40 and 60 MeV respectively.}
\end{figure}

\begin{figure}[ht]
\vskip 15truecm
\includegraphics{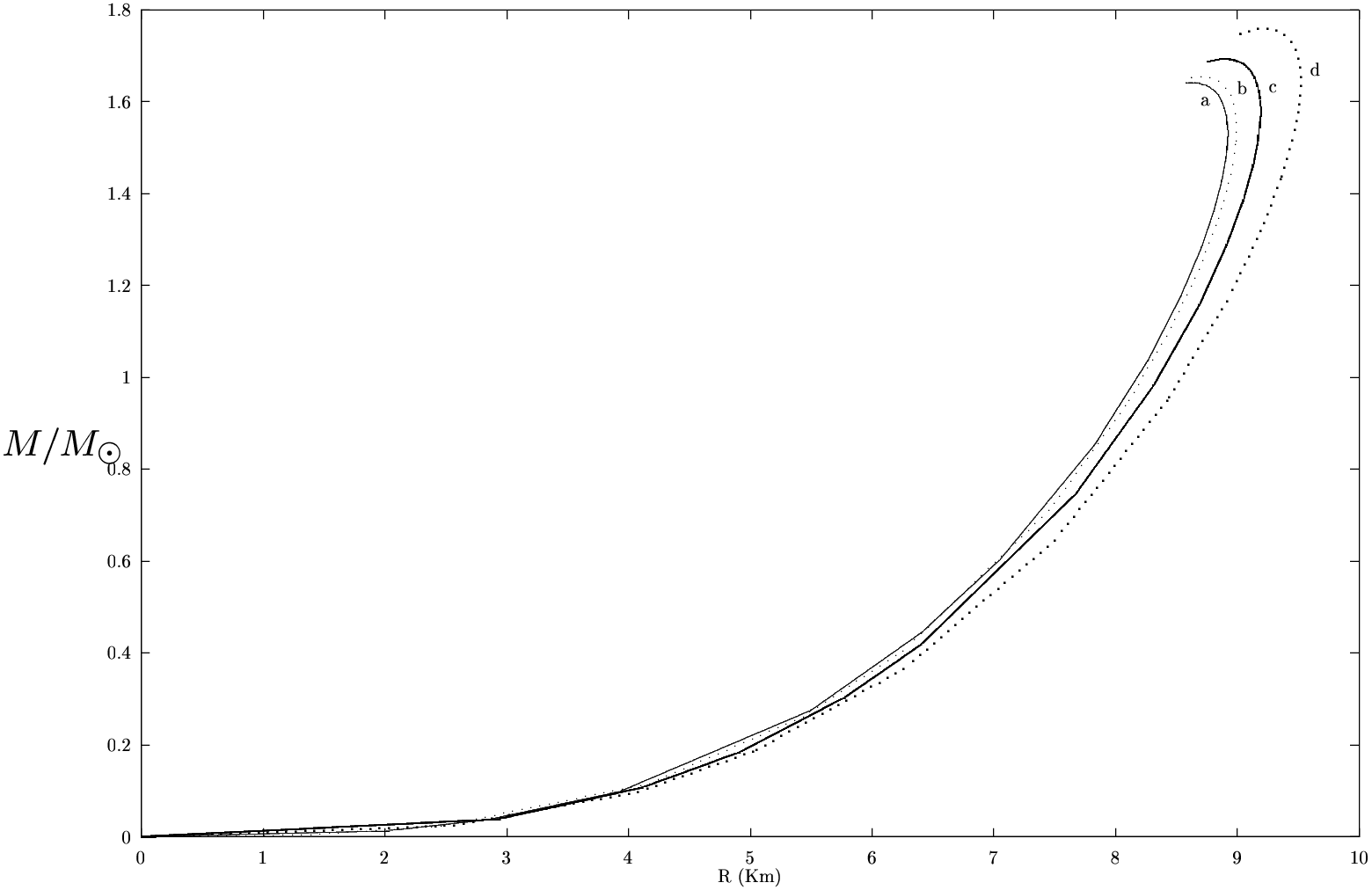}
\caption{Same as fig. 1 but for $\mu_{\nu_e}$ = 0 MeV and $\mu_{\nu_\mu}$ = 200 MeV.}
\end{figure}

\begin{figure}[ht]
\vskip 15truecm
\includegraphics{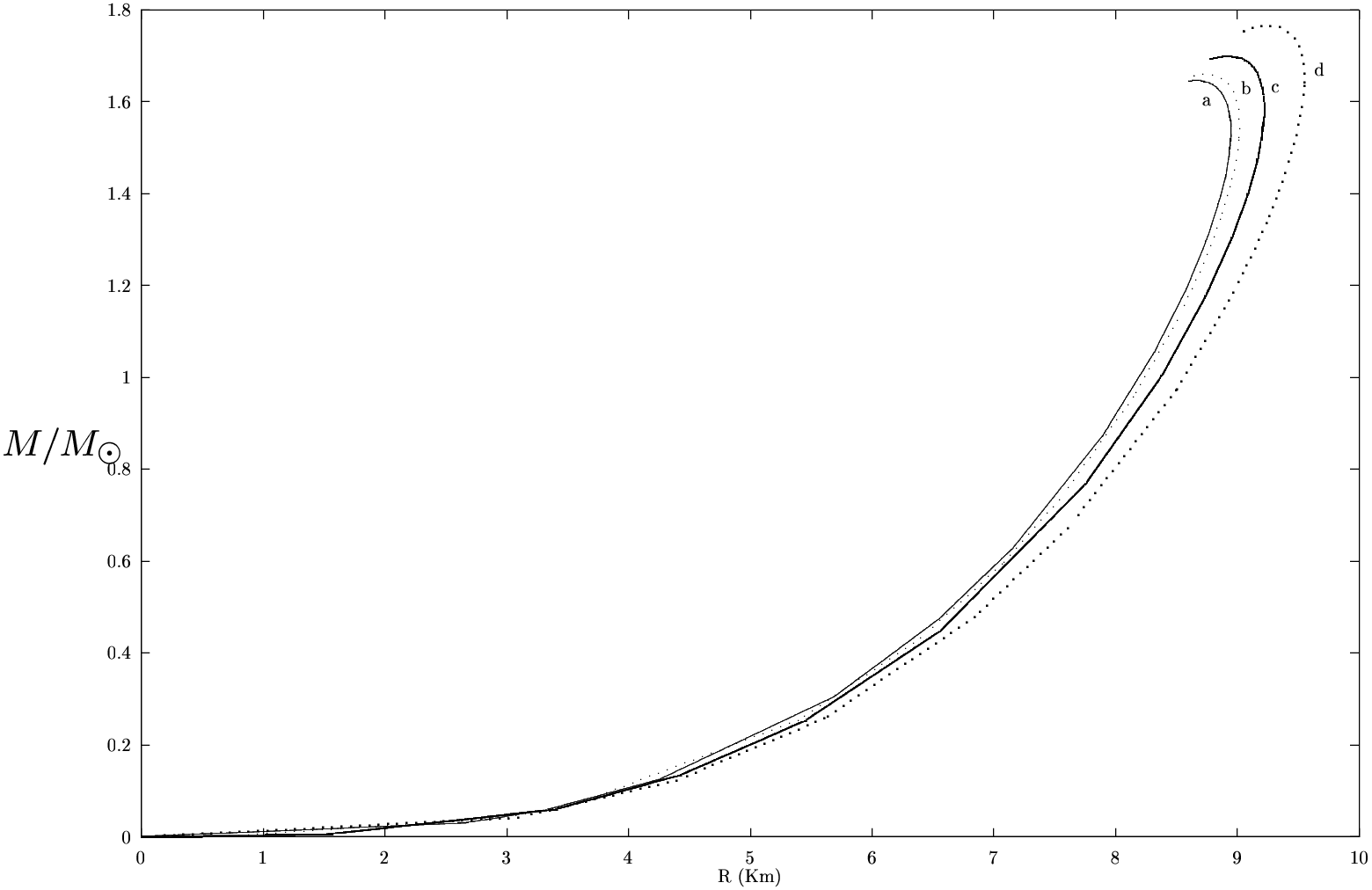}
\caption{Same as fig. 1 but for $\mu_{\nu_e}$ = 200 MeV and $\mu_{\nu_\mu}$ = 0 MeV.}
\end{figure}

\begin{figure}[ht]
\vskip 15truecm
\includegraphics{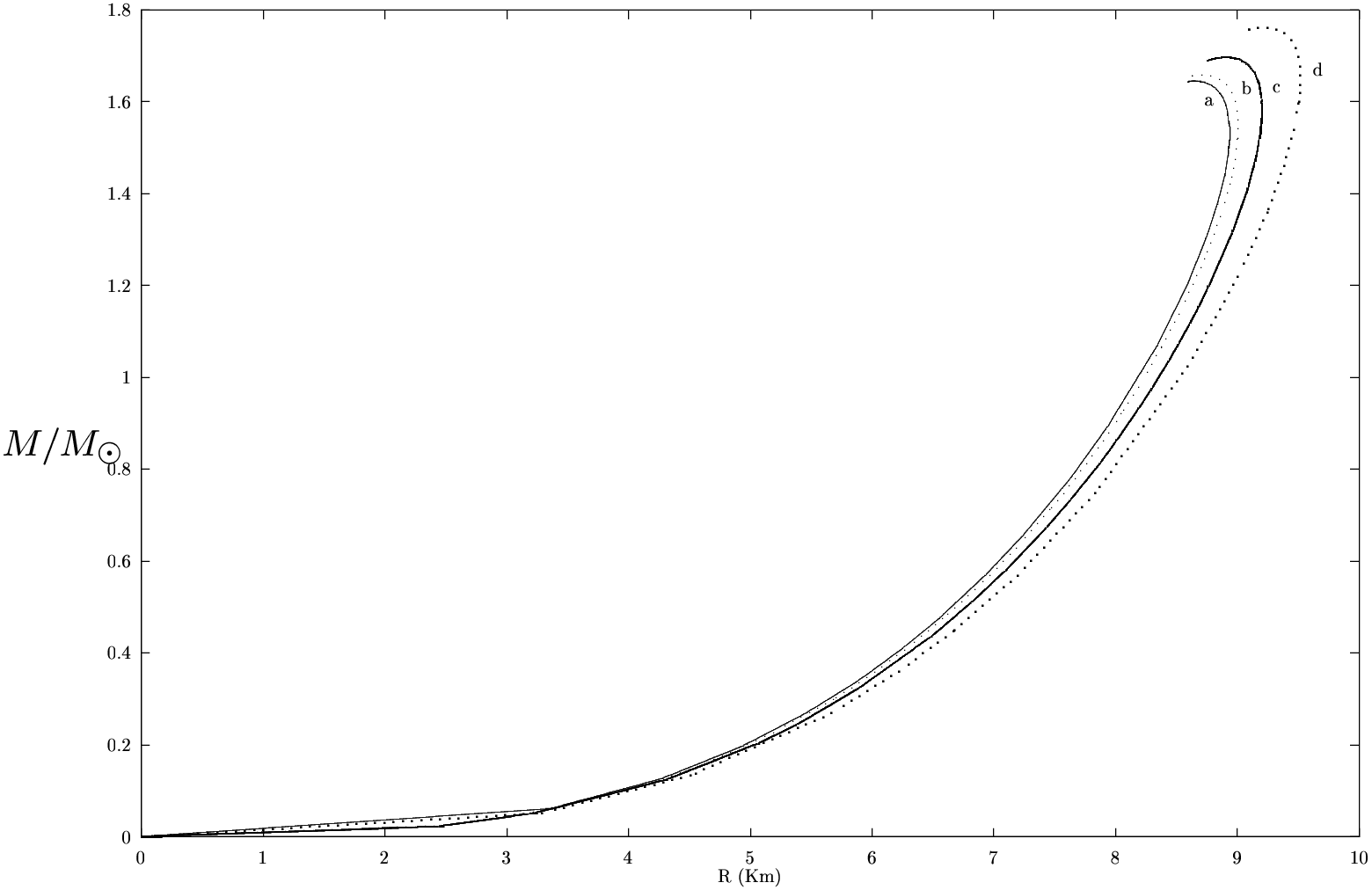}
\caption{Same as fig. 1 but for $\mu_{\nu_e}$ = $\mu_{\nu_\mu}$ = 200 MeV.}
\end{figure}

\begin{figure}[ht]
\vskip 15truecm
\includegraphics{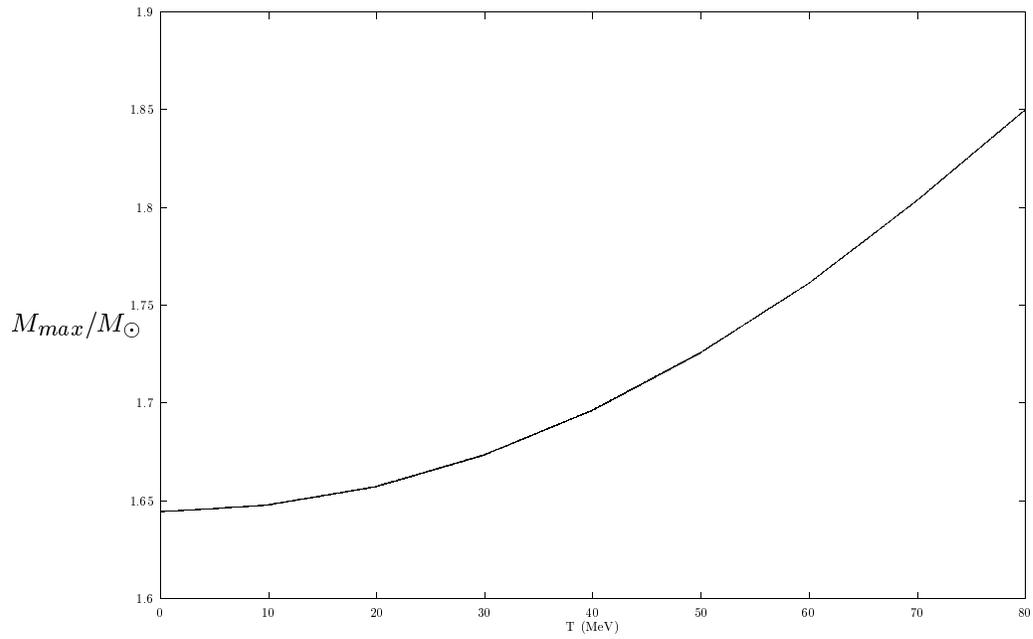}
\caption{Temperature T vs. $M_{max}/M_{\odot}$ (where $M_{max}$ is the maximum mass) for $C_o$ = 185 MeV$fm^{-3}$, $m_{so}$ = 150 MeV, $T_c$ = 170 MeV and $\mu_{\nu_e}$ = $\mu_{\nu_\mu}$ = 200 MeV.}
\end{figure}

\begin{figure}[ht]
\vskip 15truecm
\includegraphics{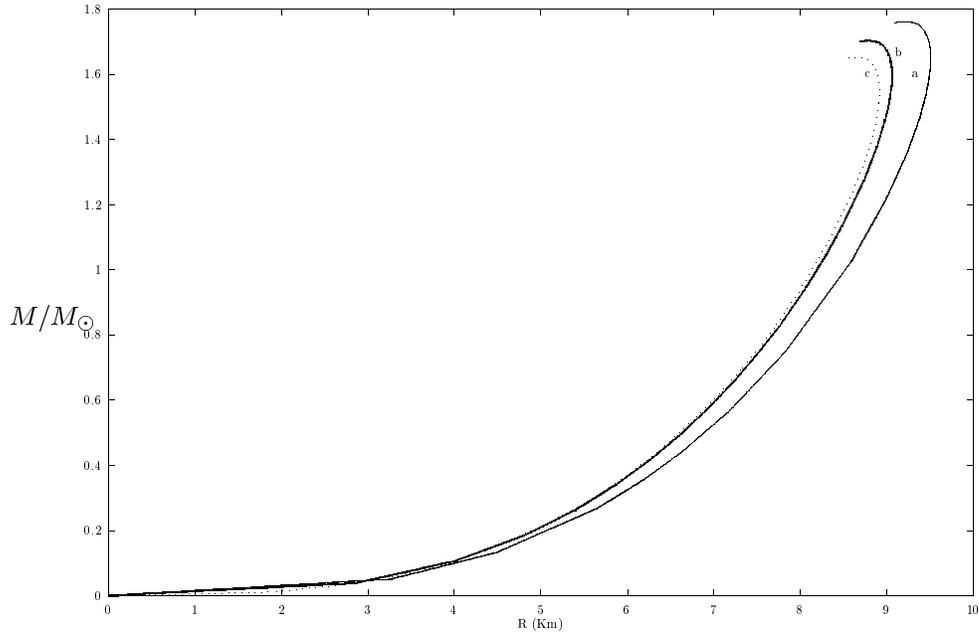}
\caption{Radius R vs. M/$M_{\odot}$ for $\mu_{\nu_e}$ = $\mu_{\nu_\mu}$ = 200 MeV and T = 60 MeV. Curves a, b correspond to $T_c$ = 170 MeV and two sets of $C_o$ $\&$ $m_{so}$ (185, 150); (210, 100) respectively. Curve c corresponds to DDQM model at C = 185 MeV$fm^{-3}$ and $m_{so}$ = 150 MeV.}
\end{figure}

\begin{figure}[ht]
\vskip 15truecm
\includegraphics{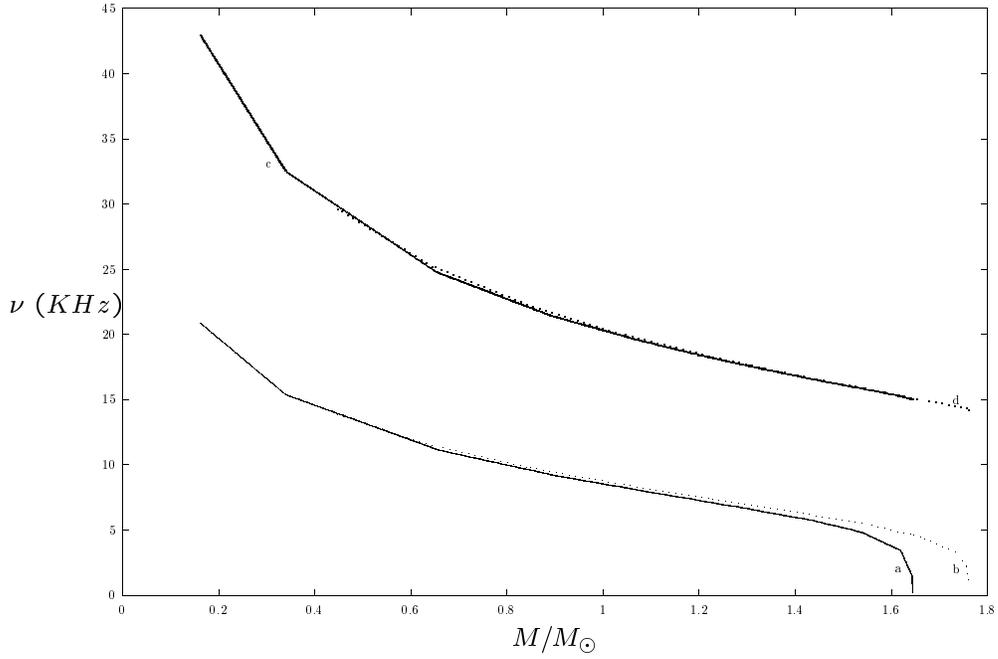}
\caption{Mass M/$M_{\odot}$ vs. frequency $\nu$ (KHz) for $C_o$ = 185 MeV$fm^{-3}$, $m_{so}$ = 150 MeV, $T_c$ = 170 MeV and $\mu_{\nu_e}$ = $\mu_{\nu_\mu}$ = 200 MeV. Curves a, b correspond to fundamental mode, n=0 at T = 0 and 60 MeV respectively. Curves c, d correspond to n=1 mode at T = 0 and 60 MeV respectively.}
\end{figure}

\begin{figure}[ht]
\vskip 15truecm
\includegraphics{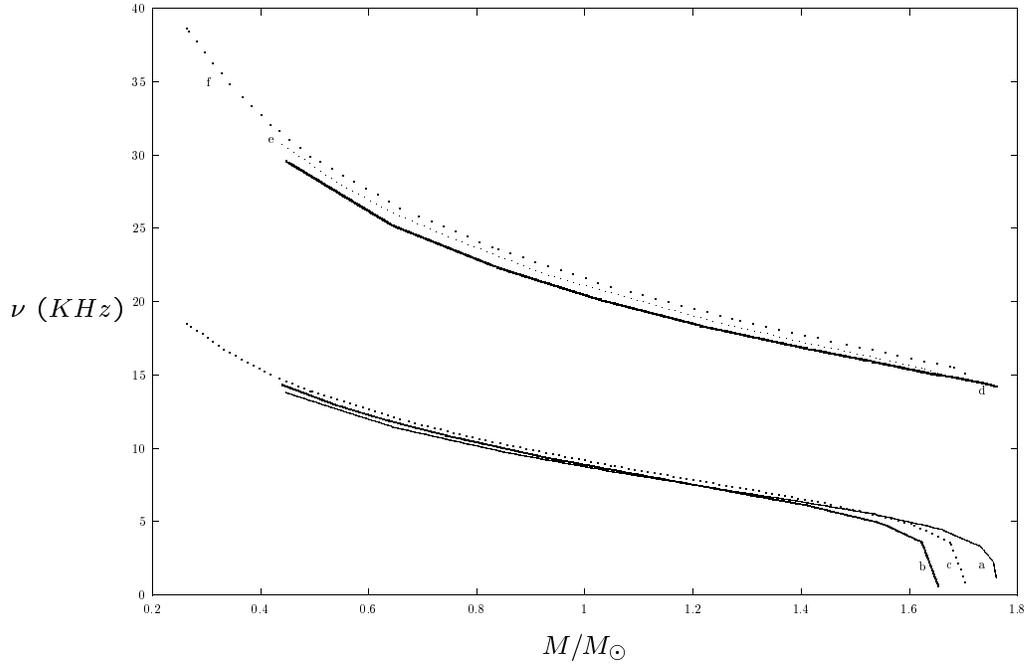}
\caption{Mass M/$M_{\odot}$ vs. frequency $\nu$ (KHz) for $\mu_{\nu_e}$ = $\mu_{\nu_\mu}$ = 200 MeV and T = 60 MeV. Curves a, c correspond to $T_c$ = 170 MeV and two sets of $C_o$ $\&$ $m_{so}$ (185, 150); (210, 100) respectively for fundamental mode. Curves b, e corresponds to corresponding DDQM model at C = 185 MeV$fm^{-3}$ and $m_{so}$ = 150 MeV.}
\end{figure}

\end{document}